\documentclass[aps,prc,letterpaper,11pt,twoside,tightenlines,nofootinbib,preprint]{revtex4}
\usepackage{graphicx}
\usepackage{epsfig,float}
\usepackage[sort&compress]{natbib}
\usepackage{subfigure}
\usepackage{amsmath}
\usepackage{amsfonts}
\usepackage{cancel}
\usepackage{lmodern,dsfont}
\usepackage{amssymb}
\usepackage{hyperref}
\begin{document}
\arraycolsep1.5pt
\newcommand{\Ima}{\textrm{Im}}
\newcommand{\Rea}{\textrm{Re}}
\newcommand{\mev}{\textrm{ MeV}}
\newcommand{\gev}{\textrm{ GeV}}
\newcommand{\dtres}{d^{\hspace{0.1mm} 3}\hspace{-0.5mm}}
\newcommand{\rts}{ \sqrt s}
\newcommand{\non}{\nonumber \\[2mm]}
\newcommand{\eps}{\epsilon}
\newcommand{\half}{\frac{1}{2}}
\newcommand{\thalf}{\textstyle \frac{1}{2}}
\newcommand{\Nmass}{M_{N}} 
\newcommand{\delmass}{M_{\Delta}} 
\newcommand{\pimass}{\mu}  
\newcommand{\rhomass}{m_\rho} 
\newcommand{\piNN}{f}      
\newcommand{\rhocoup}{g_\rho} 
\newcommand{\fpi}{f_\pi} 
\newcommand{\f}{f} 
\newcommand{\nucfld}{\psi_N} 
\newcommand{\delfld}{\psi_\Delta} 
\newcommand{\fpiNN}{f_{\pi N N}} 
\newcommand{\fpiND}{f_{\pi N \Delta}} 
\newcommand{\GMquark}{G^M_{(q)}} 
\newcommand{\vecpi}{\vec \pi}
\newcommand{\vectau}{\vec \tau}
\newcommand{\vecrho}{\vec \rho}
\newcommand{\delmu}{\partial_\mu}
\newcommand{\delMu}{\partial^\mu}
\newcommand{\nn}{\nonumber}
\newcommand{\bi}{\bibitem}
\newcommand{\vs}{\vspace{-0.20cm}}
\newcommand{\be}{\begin{equation}}
\newcommand{\ee}{\end{equation}}
\newcommand{\ba}{\begin{eqnarray}}
\newcommand{\ea}{\end{eqnarray}}
\newcommand{\ropi}{$\rho \rightarrow \pi^{0} \pi^{0}
\gamma$ }
\newcommand{\roeta}{$\rho \rightarrow \pi^{0} \eta
\gamma$ }
\newcommand{\omepi}{$\omega \rightarrow \pi^{0} \pi^{0}
\gamma$ }
\newcommand{\omeeta}{$\omega \rightarrow \pi^{0} \eta
\gamma$ }
\newcommand{\ul}{\underline}
\newcommand{\del}{\partial}
\newcommand{\rth}{\frac{1}{\sqrt{3}}}
\newcommand{\rsix}{\frac{1}{\sqrt{6}}}
\newcommand{\sq}{\sqrt}
\newcommand{\fr}{\frac}
\newcommand{\pr}{^\prime}
\newcommand{\ov}{\overline}
\newcommand{\Gm}{\Gamma}
\newcommand{\rw}{\rightarrow}
\newcommand{\rgl}{\rangle}
\newcommand{\De}{\Delta}
\newcommand{\Dp}{\Delta^+}
\newcommand{\Dm}{\Delta^-}
\newcommand{\Dz}{\Delta^0}
\newcommand{\Dpp}{\Delta^{++}}
\newcommand{\Sg}{\Sigma^*}
\newcommand{\Sp}{\Sigma^{*+}}
\newcommand{\Sm}{\Sigma^{*-}}
\newcommand{\Sz}{\Sigma^{*0}}
\newcommand{\X}{\Xi^*}
\newcommand{\Xm}{\Xi^{*-}}
\newcommand{\Xz}{\Xi^{*0}}
\newcommand{\Om}{\Omega}
\newcommand{\Omm}{\Omega^-}
\newcommand{\kp}{K^+}
\newcommand{\kz}{K^0}
\newcommand{\pip}{\pi^+}
\newcommand{\pim}{\pi^-}
\newcommand{\piz}{\pi^0}
\newcommand{\et}{\eta}
\newcommand{\kb}{\ov K}
\newcommand{\km}{K^-}
\newcommand{\kbz}{\ov K^0}
\newcommand{\ksb}{\ov {K^*}}

\def\tstrut{\vrule height2.5ex depth0pt width0pt} 
\def\jtstrut{\vrule height5ex depth0pt width0pt} 

\title{Tests on the molecular structure of  $f_2(1270)$, $f'_2(1525)$ from
$\psi (nS)$ and $\Upsilon (nS)$ decays}
\author{Lianrong Dai,$^{1,2}$  Eulogio Oset$^{2}$ \\
{\small{\it $^1$ Department of Physics, Liaoning Normal University, Dalian, 116029, China}}\\
{\small{\it $^2$Departamento de F\'{\i}sica Te\'orica and IFIC,
Centro Mixto Universidad de Valencia-CSIC,}}\\
{\small{\it Institutos de
Investigaci\'on de Paterna, Aptdo. 22085, 46071 Valencia, Spain}}\\}

\date{\today}

\begin{abstract}
Based on  previous studies that support the vector-vector molecular structure of the
$f_2(1270)$, $f'_2(1525)$, $\bar{K}^{*\,0}_2(1430)$, $f_0(1370)$ and $f_0(1710)$
resonances, we make predictions for $\psi (2S)$ decay into
$\omega(\phi) f_2(1270)$, $\omega(\phi) f'_2(1525)$, $K^{*0}(892) \bar{K}^{*\,0}_2(1430)$
and radiative decay of $\Upsilon (1S),\Upsilon (2S), \psi (2S)$ into $\gamma f_2(1270)$, $\gamma f'_2(1525)$, $\gamma f_0(1370)$, $\gamma f_0(1710)$.
Agreement with experimental data is found for three available ratios, without using free parameters,
and predictions are done for other cases.

\end{abstract}

\maketitle

\section{Introduction}
\label{Intro}

The study of $J/\psi$ decays is turning into a fruitful source of hadron
spectroscopy \cite{Shen:2008jaa,Ping:2010zz,BES:2011aa,Olsen:2012zz}.
The large mass of $J/\psi$ that allows the decay into many states, and
its precise quantum numbers, permit a selection of channels in the decay
modes where the quantum numbers can be induced with precision. It has
also turned out a good tool to study the nature of some resonances, in
particular those generated dynamically from the interaction of pairs of
more elementary hadrons. Pioneering work along this line was done in the
study of $J/\psi$ decay into $\phi$ or $\omega$ and a scalar meson
$\sigma(600)$ and $f_0(980)$ \cite{Meissner:2000bc,Roca:2004uc,Lahde:2006wr,Liu:2009ub}.
The idea is that there is a primary
$J/\psi$ decay into a vector meson ($\phi$ or $\omega$) and a pair of
pseudoscalar mesons (PP) in a scalar state. Simple arguments of SU(3)
symmetry allow to relate the couplings to the different PP channels and
then these channels are allowed to undergo rescattering and generate
resonances if this is the case, like the $\sigma(600)$ and $f_0(980)$.
Then up to a global normalization one can evaluate theoretically the
mass distributions in different channels. A follow up of this idea was
done in the study of the $J/\psi$ decay into $\phi$, $\omega$ or
$K^*(892)$ and some tensor resonances like the  $f_2(1270)$,
$f'_2(1525)$ and $K^*_2(1430)$ \cite{daizou}. The idea here is that the
$J/\psi$ decay proceeds via decay into a $\phi$, $\omega$ or $K^*(892)$
that acts as spectator in the reaction and another pair of vector mesons
in a certain invariant mass region that interact to create the
$f_2(1270)$, $f'_2(1525)$ and $K^*_2(1430)$ resonances, that, according
to \cite{gengvec}, are dynamically generated by the interaction of pairs
of vectors. These pair of vector mesons rescatter after the primary
production, producing the resonances that can be observed in the
invariant mass distributions.

  The results obtained in those former works and the good agreement with
experiment gave extra support to the idea of the $f_2(1270)$,
$f'_2(1525)$ and $K^*_2(1430)$ as being quasimolecular states of two
vectors for which other experimental support is also available
\cite{Oset:2010ab}.

  Another source of support for this idea stems from the radiative
$J/\psi$ decay into $f_2(1270)$, $f'_2(1525)$, $f_0(1370)$ and
$f_0(1710)$ \cite{hanhart}. In this case the photon is radiated from the
charmed quarks and a pair of vectors is
primary formed, leading through rescattering to the formation of these
resonances.

  The arguments used to make predictions are very simple and general and
can be easily extended to decays of $\psi(nS)$ and $\Upsilon(nS)$.
Taking advantage that some of these decays have become recently
available, we study here this extension and make the theoretical
predictions along the former lines comparing with data when available.
The agreement of the results with the data available on $\Upsilon(1S)$
to $\gamma~ f_2(1270)$ and $\gamma~ f'_2(1525)$ is good and the same
occurs with the decay of $\psi(2S)$ into $\omega~ f_2(1270)$,
$\phi~f'_2(1525)$ and $K^*(892)~K^*_2(1430)$. We also make predictions
for other related decays not yet available.

\section{Formalism for $J/\psi$ decay into  $\omega(\phi) VV \rightarrow \omega(\phi) T$}
We summarize here the formalism used in \cite{daizou} to study the decay of $J/\psi$ into
$\omega(\phi)$ and two interacting vectors that lead to the tensor state. By following the approach of
\cite{Roca:2004uc} we write the $\phi$ and $\omega$ as a combination of a singlet and an octet of
SU(3) states
\begin{eqnarray}
 \phi&=&s\bar{s}=\sqrt{\frac{1}{3}}V_1-\sqrt{\frac{2}{3}}V_8\nonumber\\
\omega&=&\frac{1}{\sqrt{2}}(u\bar{u}+d\bar{d})=\sqrt{\frac{2}{3}}V_1+\sqrt{\frac{1}{3}}V_8
\end{eqnarray}

The two extra $V'V'$ states combine to $I=0$, either with $s\bar{s}$ or $\frac{1}{\sqrt{2}}(u\bar{u}+d\bar{d})$
SU(3) structure. By taking into account that the $J/\psi$ acts as a singlet state of SU(3), we obtained a matrix element
for the $J/\psi\rightarrow\phi s\bar{s}\rightarrow \phi V'V'$ and
$J/\psi\rightarrow\phi\frac{1}{\sqrt{2}}(u\bar{u}+d\bar{d})\rightarrow\phi V'V'$ given by
\begin{eqnarray}
 \frac{1}{3}T^{(1,1)}+\frac{2}{3}T^{(8,8)}\quad\quad\mbox{and}\quad\quad
\frac{\sqrt{2}}{3}T^{(1,1)}-\frac{\sqrt{2}}{3}T^{(8,8)}
\end{eqnarray}
respectively. Similarly for the $J/\psi\rightarrow\omega s\bar{s}\rightarrow\omega V'V'$
and $J/\psi\rightarrow\omega\frac{1}{\sqrt{2}}(u\bar{u}+d\bar{d})\rightarrow\omega V'V'$ amplitudes
we obtain
\begin{eqnarray}
 \frac{\sqrt{2}}{3}T^{(1,1)}-\frac{\sqrt{2}}{3}T^{(8,8)}\quad\quad
\mbox{and}\quad\quad
\frac{2}{3}T^{(1,1)}+\frac{1}{3}T^{(8,8)}
\end{eqnarray}
respectively.  Here $T^{(1,1)}$ is the T matrix for the singlet of $\phi$ and the one of $V'V'$ giving the vacuum and $T^{(8,8)}$
the corresponding part for the octet.

Then, up to a global normalization, all decay states were given in terms of the ratio  $\nu$
\begin{eqnarray}
 \nu=\frac{T^{(1,1)}}{T^{(8,8)}}\label{nu}
\end{eqnarray}
It was also found in \cite{daizou} that in terms of $V'V'$ the $s\bar{s}$ and $\frac{1}{\sqrt{2}}(u\bar{u}+d\bar{d})$
components could be written as
\begin{eqnarray}
 s\bar{s}\rightarrow K^{*-}K^{*+}+\bar{K}^{*0}K^{*0}+\phi\phi,
\end{eqnarray}
\begin{eqnarray}
 \frac{1}{\sqrt{2}}(u\bar{u}+d\bar{d})\rightarrow
\frac{1}{\sqrt{2}}(\rho^0\rho^0+\rho^+\rho^-+\rho^-\rho^+ +\omega\omega+K^{*+}K^{*-}+K^{*0}\bar{K}^{*0}).
\end{eqnarray}

\begin{figure}[h!]
\includegraphics[width=\textwidth]{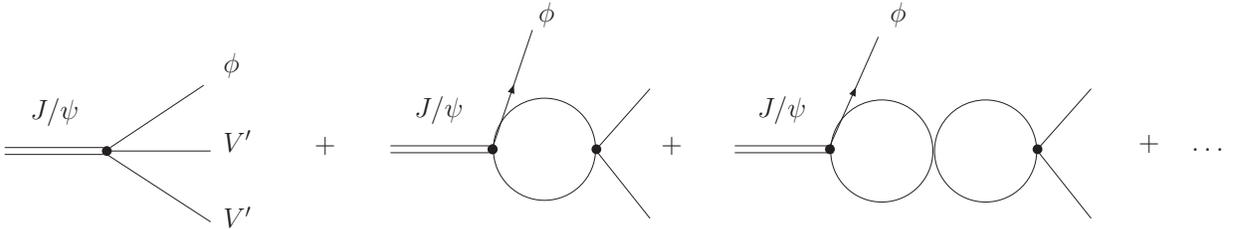}
\caption{Production mechanism  of $\phi$ and two interacting vector mesons.}\label{vertexJ}
\end{figure}

Then the production of a resonance proceeds diagrammatically as depicted in Fig. \ref{vertexJ}

\begin{figure}[h!]
\centering
\includegraphics[scale=0.8]{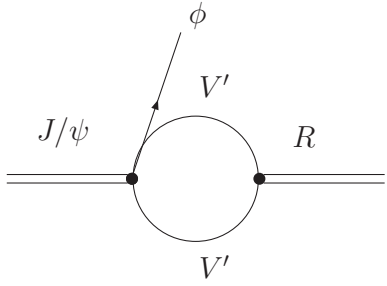}
\caption{Selection of diagrams of Fig. \ref{vertexJ} that go into resonance formation, omitting the coupling to $V^\prime V^\prime$
without interaction.}\label{JRes}
\end{figure}

Only the terms in Fig. \ref{vertexJ} where there is interaction of $V'V'$ lead to the tensor resonance and
hence this can be depicted by means of Fig. \ref{JRes}

The final transition matrix for $J/\psi\to\phi(\omega,{K^*}^0)R$,
where $R$ is the resonance studied, is given by
\begin{eqnarray}
t_{J/\psi\to\phi R}=\sum_{j}W_{j}^{(\phi)}G_{j}g_{j}\label{tJ}
\end{eqnarray}
where $W_{j}^{(\phi)}$ are weights given in \cite{daizou}, $G_{j}$ the  $V'V'$  loop functions
and $g_{j}$ the couplings of whichever resonance to the corresponding $V'V'$ channel $j$. The same can be done
for $J/\psi \rightarrow \omega R$ or $J/\psi \rightarrow {K^*}^0 R$ and all values of $\omega, G, g$ are tabulated
in \cite{daizou}.

The $J/\psi$ decay width is then given by
\begin{eqnarray}
\Gamma=\frac{1}{8\pi}\frac{1}{M^2_{J/\psi}}|t|^2q \label{Gama}
\end{eqnarray}
with $q$ the momentum of the $\phi (\omega, K^{*\,0})$ in the $J/\psi$ rest frame.

In principle the spectator $\phi (\omega , K^*)$ could interact with one of the vectors of the interacting pair that leads to the resonance considered. However, an inspection of the kinematics shows that one is far away of any of the known resonances where this new pair could have a sizeable interaction. 

Here we want to extend these ideas to the same decay channels but from the $\psi (2S)$ state.
In terms of quarks the $\psi (2S)$ state is the $c\bar{c}$ state with the same
quantum numbers as the $J/\psi$, corresponding to a radial excitations to the $2S$ state of
a $c\bar{c}$ potential. One then expects the same dynamical features as for the $J/\psi$, which stands
as the $1S$ $c\bar{c}$ state. Only two dynamical features have been taken from the $J/\psi$ concerning theses decays.
First that it is an SU(3) singlet. This certainly can be said of the  $\psi (2S)$ state too. The other dynamical feature
was related to the OZI violation and the weight going into $\phi s\bar{s}$ or $\phi\frac{1}{\sqrt{2}}(u\bar{u}+d\bar{d})$
in its decay, that we have parametrized in terms of the $\nu$ parameter of Eq. (\ref{nu}). Since
this is also a dynamical feature not related to the internal excitation of the $c\bar{c}$ quarks in the potential well, we shall
also assume that the  $\nu$ parameter  is the same for $\psi (2S)$ as for $J/\psi$. With these two reasonable assumptions we can make
predictions for the four ratios
\begin{eqnarray}
\widetilde{R_{1}}\equiv\frac{\Gamma_{\psi (2S)\to\phi f_{2}(1270)}}{\Gamma_{\psi (2S)\to\phi f^\prime_{2}(1525)}},\quad
\widetilde{R_{2}}\equiv\frac{\Gamma_{\psi (2S)\to\omega f_{2}(1270)}}{\Gamma_{\psi (2S)\to\omega f^\prime_{2}(1525)}},
\label{R2s1}
\end{eqnarray}
\begin{eqnarray}
\widetilde{R_{3}}\equiv\frac{\Gamma_{\psi (2S)\to\omega f_{2}(1270)}}{\Gamma_{\psi (2S)\to\phi f_{2}(1270)}},\quad
\widetilde{R_{4}}\equiv\frac{\Gamma_{\psi (2S)\to K^{*\,0} \bar{K}^{*\,0}_{2}(1430)}}{\Gamma_{\psi (2S)\to\omega f_{2}(1270)}}.
\label{R2s2}
\end{eqnarray}
The values of these ratios with their theoretical uncertainties for the $J/\psi$ decay are given in Table 7 of \cite{daizou}
 which we reproduce below in Table \ref{Table1:R}. In order to get the ratios for $\psi (2S)$ we simply have to change the momenta that enter the formula of the width, Eq. (\ref{Gama}), since now the mass of $\psi (2S)$ is different from the one of $J/\psi$.
When this is corrected we find the values of Table \ref{R2s},
 \begin{table}[h]
\caption{Comparison between the experimental and the theoretical results of $J/\psi$ decays taken from \cite{daizou}.
The ratios $R_{i}$ are defined as in Eq. (\ref{R2s1}), (\ref{R2s2}) but for $J/\psi$ instead of $\psi (2S)$.  }
\begin{center}
\begin{tabular}{ccc}
\hline\hline
&Experiment&Theory\\
\hline
\\
$R_{1}$&0.22 - 0.47 $(0.33^{+0.14}_{-0.11})$&0.13 - 0.61 $(0.28^{+0.33}_{-0.15})$\\[1.5ex]
$R_{2}$&12.33 - 49.00 $(21.50^{+27.50}_{-9.17})$&2.92 - 13.58 $(5.88^{+7.70}_{-2.96})$\\[1.5ex]
$R_{3}$&11.21 - 23.08 $(15.85^{+7.23}_{-4.65})$&6.18 - 19.15 $(10.63^{+8.52}_{-4.45})$\\[1.5ex]
$R_{4}$&0.55 - 0.89 $(0.70^{+0.19}_{-0.15})$&0.83 - 2.10 $(1.33^{+0.77}_{-0.50})$\\[1.5ex]
\hline
\end{tabular}
\label{Table1:R}
\end{center}
\end{table}
where we compare with experimental values when possible. There is only the ratio $\widetilde{R_{4}}$ directly available, but one experimental ratio,
$\Gamma_{\psi (2S)\to\omega f_{2}(1270)}/\Gamma_{\psi (2S)\to\phi f^\prime_{2}(1525)}$, can be obtained from the product of $\widetilde{R_{1}}\cdot\widetilde{R_{3}}$ and we also show the
results in Table \ref{R2s}. The theoretical errors are evaluated as in \cite{daizou}. Hence the relative errors of each ratio $\widetilde{R_{i}}$
are the same as for $R_{i}$  of Table \ref{Table1:R}. The relative errors for the theoretical $\widetilde{R_{1}}\cdot\widetilde{R_{3}}$ corresponding
to $\Gamma_{\psi (2S)\to\omega f_{2}(1270)}/\Gamma_{\psi (2S)\to\phi f^\prime_{2}(1525)}$ are like those of $R_{1}$, where they are biggest.

As we can see, the agreement of the theory with the experimental result is good, although one has large uncertainties in both cases. Note, however, that even then, the agreement is not third aim, as one can see in Table \ref{R2s}, there are some times difference of a factor 40 between some of the ratios.

In Table \ref{R2s} we also make predictions for other ratios to allow comparison with future  experiments that could be stimulated by the present  findings.

\begin{table}[h]
\caption{Comparison between the experimental and the theoretical results of $\psi (2S)$ decays.
$\widetilde{R_{1}}\cdot\widetilde{R_{3}}$ provides the ratio $\Gamma_{\psi (2S)\to\omega f_{2}(1270)}/\Gamma_{\psi (2S)\to\phi f^\prime_{2}(1525)}$.
Experimental results for $\widetilde{R_{4}}$ and $\widetilde{R_{1}}\cdot\widetilde{R_{3}}$  from \cite{pdg,bes04,bes98}.}
\begin{center}
\begin{tabular}{ccc}
\hline\hline
&Experiment&Theory\\
\hline
\\
$\widetilde{R_{1}}$& & 0.12-0.56 $(0.26^{+0.30}_{-0.14})$\\[1.5ex]
$\widetilde{R_{2}}$& &2.76-12.82$(5.55^{+7.27}_{-2.79})$\\[1.5ex]
$\widetilde{R_{3}}$& &5.91-18.30$(10.16^{+8.14}_{-4.25})$\\[1.5ex]
$\widetilde{R_{4}}$&0.54-1.33 $(0.86^{+0.47}_{-0.32})~~~$ & 0.88-2.22 $(1.41^{+0.82}_{-0.53})$ \\[1.5ex]
$\widetilde{R_{1}}\cdot\widetilde{R_{3}}~~~$& 3.0-9.3 $(5.0^{+4.3}_{-2.0})~~~$ & 1.24-5.74 $(2.64^{+3.1}_{-1.4})$\\[1.5ex]
\hline
\end{tabular}
\label{R2s}
\end{center}
\end{table}

\section{$\gamma$ radiative decay into  $\gamma VV \rightarrow \gamma T$}
\begin{figure}[htpb]
\begin{center}
\includegraphics[scale=0.6]{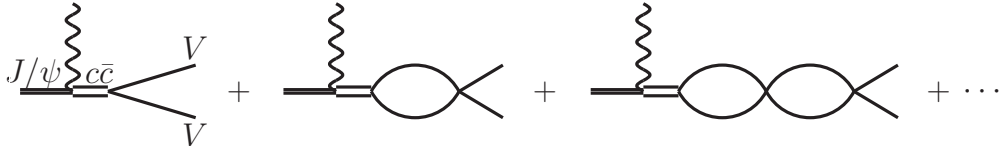}
\caption{Schematic representation of $J/\psi$ decay into a photon and one dynamically generated resonance.} \label{f3}
\end{center}
\end{figure}

Another successful test on the vector-vector nature of the $f_2(1270)$ and $f'_2(1525)$  resonances
was done in \cite{hanhart} looking at the radiative decay of $J/\psi$  into $\gamma T$, with T any
of these two tensor resonances. In \cite{hanhart} it was justified that the photon was radiated from the
initial $c\bar{c}$ state and a remaining  $c\bar{c}$ decomposed into a pair of vector mesons which
upon rescattering gave rise to the tensor resonances. The mechanism is deplicted in Fig. \ref{f3}

The transition amplitude is given by
\begin{eqnarray}\label{eq:mat}
 t_{J/\psi\rightarrow \gamma \mathrm{R}}=\sum_j \widetilde{w_j} G_j g_j \ .
\end{eqnarray}
and the weights $\widetilde{w_j}$ are obtained in \cite{hanhart} and given by
\begin{eqnarray}\label{eq:cg}
\widetilde{ w_i}=a~\left\{\begin{array}{ll}
             -\sqrt{\frac{3}{2}}&\quad\mbox{for $\rho\rho$}\\
             -\sqrt{2}&\quad\mbox{for $K^*\bar{K}^*$}\\
             \frac{1}{\sqrt{2}} &\quad \mbox{for $\omega\omega$}\\
             \frac{1}{\sqrt{2}} &\quad\mbox{for $\phi\phi$}
            \end{array}
\right. .
\end{eqnarray}
where a is a normarlization constant, inoperative in the ratios, and
$G_j$, $g_j$ are again the loop functions of the intermediate $V'V'$ states and
the couplings of the resonance to these $V'V'$ states. All the needed quantities for the evaluation
of the amplitude are given in Table 1 of \cite{hanhart}. The same theoretical framework allowed to evaluate
the $J/\psi$ radiative decay into the scalar meson $f_0(1370)$ and $f_0(1710)$  which are also obtained
from the interaction of $V'V'$, mostly $\rho\rho$ and $K^{*0} \bar{K}^{*\,0}$ respectively. The decay width is
given again by Eq. (\ref{Gama}) where $q$ is now  the momentum of the photon in the $J/\psi$  rest frame.

\begin{figure}[b]
\begin{center}
\includegraphics[scale=0.6]{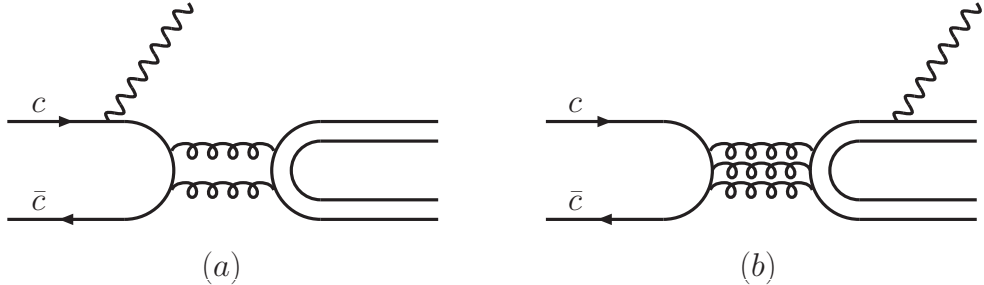}
\caption{Two mechanisms of the $J/\psi$ radiative decays.}
\label{f4}
\end{center}
\end{figure}

In the present case we want to extend these results to the decay of the $\Upsilon$. The only dynamical assumption made in
\cite{hanhart} was that the photon was radiated from the $c\bar{c}$ state, not from the final $V'V'$ state.
The argument was based on the dominance
of the diagram of Fig.\ref{f4}(a) over the one of Fig.\ref{f4}(b), which require two, versus three, gluon exchange
as discussed in \cite{Kopke:1988cs}. On the other hand the $c\bar{c}$ state was assumed to be a singlet of SU(3)
like in the case of the former section. Both assumptions hold equally here for the case of the $b\bar{b}$  in  $\Upsilon$ decay and,
hence the only difference  in the results stems from an overall normalization, which disappears when  ratios are made,
and the momenta $q$ in the formula of the width, since now the $\Upsilon$ mass is different to the one of the $J/\psi$. When this is taken into account, we evaluate the ratios that we define below in Eq. (\ref{eq:RTRS}):

\begin{eqnarray}\label{eq:RTRS}
\begin{array}{ll}
             R_T=\frac{\Gamma_\mathrm{J/\psi\rightarrow \gamma f_2(1270)}}{\Gamma_{J/\psi\rightarrow
\gamma f'_2(1525)}}; & R_S=\frac{\Gamma_{J/\psi\rightarrow \gamma f_0(1370)}}{\Gamma_{J/\psi\rightarrow\gamma
f_0(1710)}},\\
\widetilde{R_T}= \frac{\Gamma_\mathrm{\Upsilon(1S)\rightarrow \gamma f_2(1270)}}{\Gamma_{\Upsilon(1S)\rightarrow
\gamma f'_2(1525)}};   & \widetilde{R_S}=\frac{\Gamma_{\Upsilon(1S)\rightarrow \gamma f_0(1370)}}{\Gamma_{\Upsilon(1S)\rightarrow\gamma
f_0(1710)}},\\
   \widehat{R_T}= \frac{\Gamma_\mathrm{\psi(2S)\rightarrow \gamma f_2(1270)}}{\Gamma_{\psi (2S)\rightarrow
\gamma f'_2(1525)}}; &\widehat{R_S}=\frac{\Gamma_{\psi (2S)\rightarrow \gamma f_0(1370)}}{\Gamma_{\psi(2S)\rightarrow\gamma
f_0(1710)}}, \\
 \overline{R_T}=\frac{\Gamma_\mathrm{\Upsilon(2S)\rightarrow \gamma f_2(1270)}}{\Gamma_{\Upsilon(2S)\rightarrow
\gamma f'_2(1525)}}; &\overline{R_S}=\frac{\Gamma_{\Upsilon(2S)\rightarrow \gamma f_0(1370)}}{\Gamma_{\Upsilon(2S)\rightarrow\gamma
f_0(1710)}}. \\
            \end{array}
\end{eqnarray}

The results that we obtain are summarized in Table \ref{ratio}. As we can see, the agreement obtained with the data of \cite{pdg}, for $\widetilde{R_T}$, (for $\Upsilon (1S)$ decay), is good within theoretical and experimental uncertainties. In Table III, we also make predictions for the radiative decay into the scalar states and for the
 $\psi (2S)$ and $\Upsilon (2S)$ states, for comparison with future experiments.

\begin{table*}[htpb]
      \renewcommand{\arraystretch}{1.5}
     \setlength{\tabcolsep}{0.3cm}
\caption{Ratios of the molecular model and in comparison with data \cite{CLEO1,CLEO2,CLEO3}.
\label{ratio}}
\begin{center}
\begin{tabular}{c|cc}\hline\hline
 & Molecular picture &  Data\\\hline
$R_T~(J/\psi)$ & $2\pm 1 $~\cite{hanhart} & $3.18^{+0.58}_{-0.64}$ \\
$R_S~(J/\psi)$ & $1.2\pm 0.3$ & \\
\hline
 $\widetilde{R_T}~(\Upsilon (1S))$ & $1.84\pm 0.92$ & $2.66^{+1.13}_{-0.70}$ \\
$\widetilde{R_S}~(\Upsilon (1S))$ & $1.05\pm 0.26$  & \\
\hline
$\widehat{R_T}~(\psi (2S))$ & $1.94 \pm 0.97$ & \\
$\widehat{R_S}~(\psi (2S))$ & $1.14\pm 0.28$& \\
\hline
$\overline{R_T}~(\Upsilon (2S))$ & $ 1.83\pm 0.92$ &  \\
$\overline{R_S}~(\Upsilon (2S))$ & $1.05\pm 0.26$ & \\
\hline\hline
\end{tabular}
\end{center}
\end{table*}

It should be noted that a simple quark model studied in  \cite{hanhart} gave ratios of the order of 2.2 for $R_T~(J/\psi)$,  which corrected by phase space would be now 2.0 for $\widetilde{R_T}~(\Upsilon (1S))$, a ratio compatible with data. The agreement with data of the molecular picture has to be
put in addition to other magnitudes, as those of section II to discriminate among models. Yet, a striking disagreement with the data for the radiative
decay would serve to invalidate the model. This stresses  the value of the ratios discussed here. Yet, it was also observed in \cite{hanhart} that the ratio $R_S/R_T$  had small theoretical errors and was about one half the  value of the quark model. So, this ratio is discriminative by itself and should
motivate the search of radiative decays in the scalar sector.

\section{Conclusion}
We have extended a test on the molecular nature of the $f_2(1270)$, $f'_2(1525)$ and $K^*_2(1430)$ resonances, using the decay of $J/\psi$ into $\phi (\omega)$ and any of the  $f_2(1270)$, $f'_2(1525)$ resonances, or $K^*(892)$ and $K^*_2(1430)$, to the same decays from the $\psi(2S)$ state. The theory only makes use of the fact that both $J/\psi$ and $\psi(2S)$ are singlets of SU(3) and there is a dynamical factor for the OZI violation into the strange and non strange sectors, the $\nu$ parameter. The needed modifications due to kinematics have been done and results for the decays of the  $\psi(2S)$ are reported that are in agreement with experiment for the available data. Predictions are also done for ratios not yet measured.

   On the other hand we also made the test of consistency on the nature of these states for the radiative decay of heavy quarkonium states, extending the test done for $J/\psi$ decay into $\gamma ~f_2(1270)$ and $\gamma ~f'_2(1525)$ to
the decay of the $\Upsilon(1S)$ into these channels. In this case the common assumptions done is the SU(3) singlet nature of the $J/\psi$ and the $\Upsilon(1S)$ and the fact that QCD dynamics strongly favors the radiation of the photon from the quarkonium state instead of being from the final state of two vector mesons. Once again, the agreement of the results with the experimental ratio for the decay of the $\Upsilon(1S)$ into $\gamma ~f_2(1270)$ and $\gamma ~f'_2(1525)$ is good, and we make predictions for the decays into  $\gamma ~f_0(1370)$ and $\gamma ~f_0(1710)$ to be compared with future experiments. We also make predictions for these radiative decays from the $\psi (2S)$ and $\Upsilon(2S)$ states.
The work done here shows the power of using decays of heavy quarkonium into states which are dynamically generated. It puts constraints on the ratios of some of the decay channels that, if fulfilled, give us more confidence in the molecular picture emerging for these states from a dynamical study of the interaction of pairs of hadrons. Future experiments, determining gradually the ratios evaluated for which there are not yet  experimental data,  will set further tests for the structure of these resonances.

\section*{Acknowledgments}
We would like to thank Feng Kun Guo for valuable comments. This work is partly  supported by National Natural Science
Foundation of China (10975068) and the Natural Science Foundation of the Liaoning Scientific Committee.
This work is also partly supported by the Spanish Ministerio de Economiay
Competitividad and European FEDER funds under the contract number
FIS2011-28853-C02-01, and the Generalitat Valenciana in the program
Prometeo, 2009/090. We acknowledge the support of the European
Community-Research Infrastructure Integrating Activity
Study of Strongly Interacting Matter (acronym HadronPhysics3, Grant
Agreement no. 283286) under the Seventh Framework Programme of EU.

\end{document}